\documentclass[%
 reprint,
 amsmath,amssymb,
 aps,
]{revtex4-2}

\usepackage{graphicx}
\usepackage{dcolumn}
\usepackage{bm}

\begin{document}


\title{Mesoscopic Transport of Quantum Anomalous Hall Effect in Sub-Micron Size Regime}

\author{Gang Qiu$^{1*}$}
\author{Peng Zhang$^1$}
\author{Peng Deng$^1$}
\author{Su Kong Chong$^1$}
\author{Lixuan Tai$^1$}
\author{Christopher Eckberg$^{2,3,4}$}
\author{Kang L. Wang$^{1,5}$}
 \email{Corresponding authors: Gang Qiu (gqiu@g.ucla.edu), Kang L. Wang (wang@ee.ucla.edu)}
\affiliation{$^1$Department of Electrical and Computer Engineering, University of California Los Angeles, 
Los Angeles, CA, 90095, USA}
\affiliation{$^2$Fibertek Inc., Herndon, VA, 20171, USA}
\affiliation{$^3$US Army Research Laboratory, Adelphi, MD, 20783, USA
}
\affiliation{$^4$US Army Research Laboratory, Playa Vista, CA, 90094, USA
}
\affiliation{
$^5$Department of Physics and Astronomy, University of California Los Angeles, 
Los Angeles, CA, 90095, USA
}

\begin{abstract}
  The quantum anomalous Hall (QAH) effect has been demonstrated in two-dimensional topological insulator systems incorporated with ferromagnetism. However, a comprehensive understanding of mesoscopic transport in sub-micron QAH devices has yet been established. Here we fabricated miniaturized QAH devices with channel widths down to 600 nm, where the QAH features are still preserved. A back-scattering channel is formed in narrow QAH devices through percolative hopping between 2D compressible puddles. Large resistance fluctuations are observed in narrow devices near the coercive field, which is associated with collective interference between intersecting paths along domain walls when the device geometry is smaller than the phase coherence length  $L_\phi$. Through measurement of size-dependent breakdown current, we confirmed that the chiral edge states are confined at the physical boundary with its width on the order of Fermi wavelength.

\end{abstract}

\maketitle


\section{\label{sec:level1}Introduction}
The quantum anomalous Hall (QAH) effect arises when long-range ferromagnetic order was introduced into a two-dimensional (2D) topological insulator, which opens an exchange gap on the surface, and chiral edge states are formed due to broken inversion symmetry. The topologically protected chiral edge states are immune to back-scattering and thus can support dissipation-less current flow in the longitudinal direction, whereas the Hall resistance is quantized into h/e$^2$. Experimentally, the QAH effect was first demonstrated by introducing properly engineered transition metal magnetic dopants (Cr or V) into topological insulator thin films  \cite{Cui-Zu2013,Kou2014,Chang2015,Checkelsky2014}. Recently, the QAH effect has also emerged in intrinsic 2D van der Waals material-based systems, including MnBi$_2$Te$_4$  \cite{Yujun2020}, and moiré patterns in twisted bi-layer graphene  \cite{Serlin2020} and MoTe$_2$/WSe$_2$ heterostructures  \cite{Li2021}. 

Because of the zero-loss current carrying capability, QAH insulators are deemed as a promising platform to design novel quantum devices that operate at low temperatures to process quantum information  \cite{Mahoney2017}. When coupled with an s-wave superconductor, proximitized topological superconductivity can also be introduced in QAH hetero-structures for topological routes towards fault-tolerant quantum computing applications\cite{Wang2015,He2017}. In regard to implementing quantum devices, QAH systems have a unique advantage over quantum Hall (QH) systems, which require a large external magnetic field to drive two-dimensional electron gases into Landau quantization. For potential scalable manufacturing of quantum devices or topological qubits, Cr-doped (BiSe)$_2$Te$_3$ stands out as the most suitable platform among available QAH candidates because of its matured massive production route through molecular beam epitaxy (MBE). So far, the experimental efforts have been focusing on macroscopic QAH devices where the chiral edge modes are simply treated as a perfect 1D conductor at the physical boundary of the QAH mesa. When leveraging the advantageous QAH effect for quantum nano-device applications, a key question to be addressed is where the chiral edge states are physically located, and how wide the chiral edge channel is. Although the chiral edge states have been directly visualized through microwave scanning microscopy  \cite{Allen2019}, the several $\mu$m wide conductive edge should not be interpreted as the edge channel width but most likely limited by the resolution of the instrumentation. In this paper, we report a comprehensive investigation of the transport behavior in miniaturized QAH devices in the sub-micron size regime. A mesoscopic picture of chiral edge states is presented, which is in sharp contrast to the QH edge states. Our work provides an assessment for a geometrical size limit above which the QAH effect can survive and offers an insight for physical modeling and analysis of chiral edge modes in QAH nano-devices.

\section{\label{sec:level1}Results and discussion}
Magnetic topological insulator (MTI) films were prepared by MBE growth following previously reported recipes \cite{Mogi2015}. Six quintuple layers of Cr$_{0.12}$(Bi$_{0.26}$Sb$_{0.62}$)$_2$Te$_3$ film was epitaxially deposited onto GaAs (111) substrates. The samples were first patterned into a 10-$\mu$m wide Hall bar using photolithography, then the narrow channels were further defined with electron beam lithography. Detailed description of material preparation, device fabrication and measurement techniques are provided in Supplementary Note 1. An optical image and a false-colored scanning electron microscopy image of representative devices are shown in Fig. 1a and 1b. 
\begin{figure}
\includegraphics[width=\linewidth]{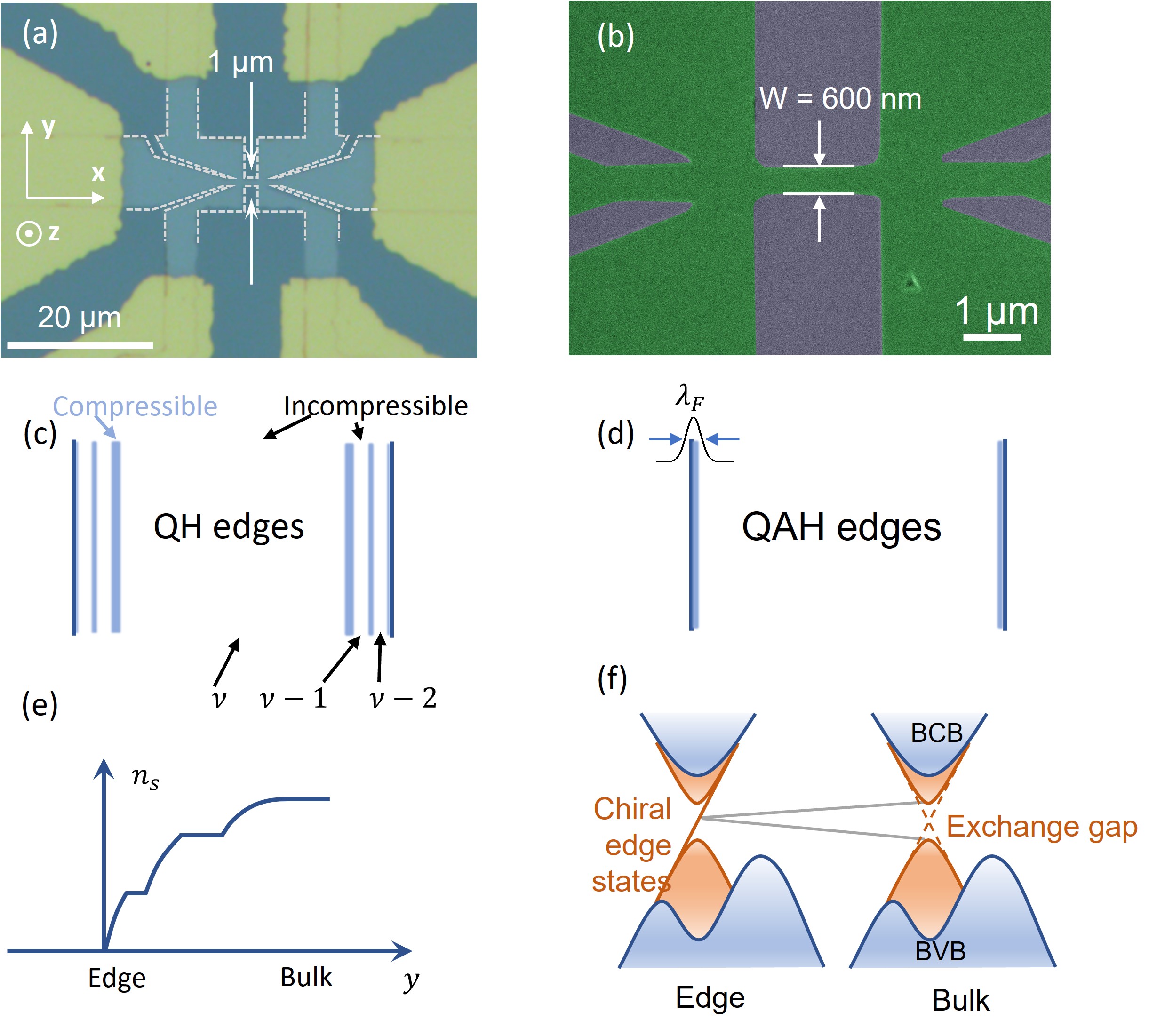}
\caption{\label{fig:epsart}  Sub-micron size quantum anomalous Hall (QAH) device for studying chiral edge modes. (a) An optical image of a 1-$\mu$m-wide Hall bar device. The channel region was first patterned with photolithography and then defined by e-beam lithography. The outline of the active channel region is marked by dashed lines. (b) A false-colored SEM image of a 0.6-$\mu$m-wide Hall bar device. The green region highlights the QAH films. (c) Schematics of bulk-boundary correspondence of a quantum Hall (QH) system where edge channels are composited of alternating compressible and incompressible edge states. (d) In a QAH system, edge channels are confined to the physical boundary. (e) Carrier density distributions of from the QH edge to the bulk. (f) Topological band evolution from edge to bulk in a QAH insulator.}
\end{figure}
To reveal size-dependent transport behaviors, a set of four devices were fabricated on the same MTI substrate to minimize sample-to-sample variations. The channel widths of the Hall-bar devices presented in the main manuscript are 5 $\mu$m, 3 $\mu$m, 1.5 $\mu$m, and 0.6 $\mu$m, respectively. Additional sets of data from a different MBE growth batch are also available in Supplementary Note 2 and Fig. S1-S3 with a similar size-dependent trend. The longitudinal resistivity $\rho_{xx}$ and Hall resistivity $\rho_{yx}$ of all four devices measured at 100 mK are presented in Fig. 2a and 2b, respectively. \begin{figure}
\includegraphics[width=\linewidth]{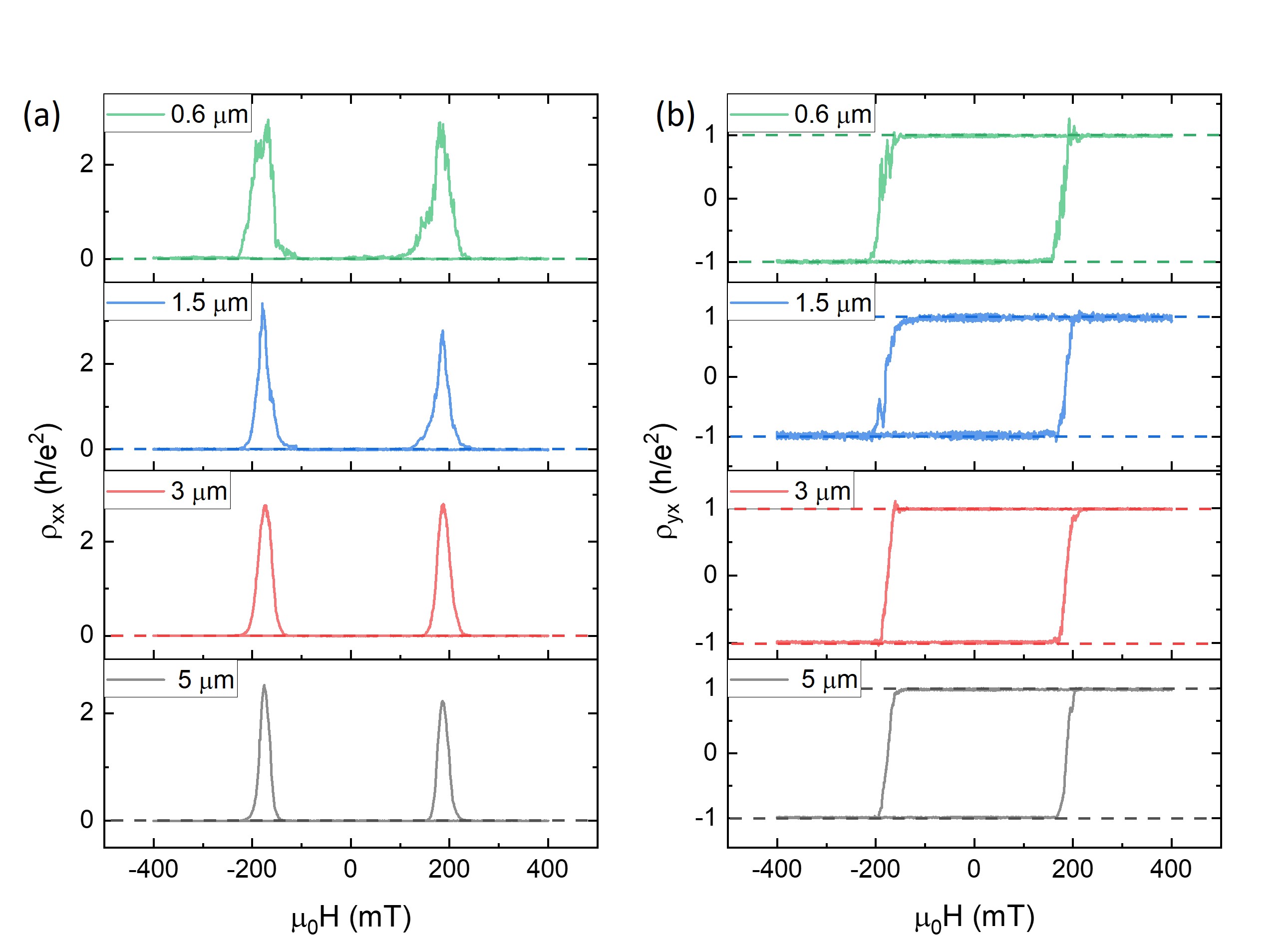}
\caption{\label{fig:epsart} Magnetic loops of sub-micron QAH devices with various sizes. (a) Longitudinal resistivity $\rho_{xx}$ and (b) transverse resistivity $\rho_{xy}$ in four devices with a channel width of 0.6, 1.5, 3, 5 $\mu$m, respectively. All data are acquired at a base temperature of 100 mK.}
\end{figure}
Here, low AC current excitation of either 0.5 nA or 1 nA was injected in all Hall bar devices unless specified otherwise, to avoid the current heating effect and transverse electric field breakdown (See detailed discussion later in Fig. 4). Depending on the magnetization directions, the transverse resistivity $\rho_{xy}$ is fully quantized at ±$h/e^2$, which is originated from integer topological invariant \textit{i.e.}, Chern number C=±1 of the system. The longitudinal resistance $\rho_{xx}$ of the narrow Hall bar also resembles its macroscopic counterpart, with a vanishing $\rho_{xx}$ when in quantized states, and a resistance peak near the coercive field $H_c$, corresponding to trivial insulating states during magnetic domain switching. The longitudinal conductivity $\sigma_{xx}$ and transverse conductivity $\sigma_{xy}$ are converted through the tensor matrix and plotted in Fig. S4 (see discussion in Supplementary Note 3), and the signatures of QAH states is again confirmed. Therefore we conclude that the general features of the QAH insulators are preserved in narrow channels down to 600 nm. 

The edge channel carrier distribution in QAH nano-devices are in sharp contrast to that of a QH system, despite that the two classes of systems share a similar topology and bulk-boundary correspondence (as illustrated in Fig. 1c and 1d). In QH states, 2D electron gases are localized under Landau quantization, and the carrier density are spatially depopulated from sheet density $n_s$ in the film to 0 at the physical edge (see Fig. 1f) across a depletion width, which is typically on the order of 1 $\mu$m  \cite{Weis2011,Patlatiuk2018}. Under the Thomas-Fermi approximation, the chiral edge channels of QH states are spatially distributed in the \textit{y}-direction in the form of alternating compressible and incompressible strips within the depletion region (see Fig. 1c) \cite{Weis2011,Yacoby2001}. Hence, two-dimensional electron gas systems will deviate from quantum Hall states when the width of the channel is reduced to the depletion width  \cite{Panos2014,Haremski2020,Eaves1986}, as the innermost incompressible strips on the opposite edges start to merge to provide a back-scattering channel. On the other hand, the incompressible bulk states in QAH insulators are gapped by magnetic exchange interaction and top-bottom surface hybridization, thus there are no carriers in the bulk nor a depletion region on the edge in QAH systems, as shown in Fig. 1d. As a result, chiral edge states are tightly confined close to the physical boundary of the films, with the wavefunction only extended by the order of Fermi wavelength  \cite{Doh2013}, \textit{i.e.}, the de Broglie wavelength of chiral edge states near the Fermi energy (typically several nm, see Fig. 1d). In this sense, the QAH effect are more robust against size-induced breakdown and can in principle survive in submicron devices.

In the next section, we will discuss several other transport features that emerge in narrow QAH devices due to the finite size effect.

The remnant resistance in $\rho_{xx}$ during field cooling is shown in an Arrhenius plot in Fig. 3a. 
\begin{figure}
\includegraphics[width=\linewidth]{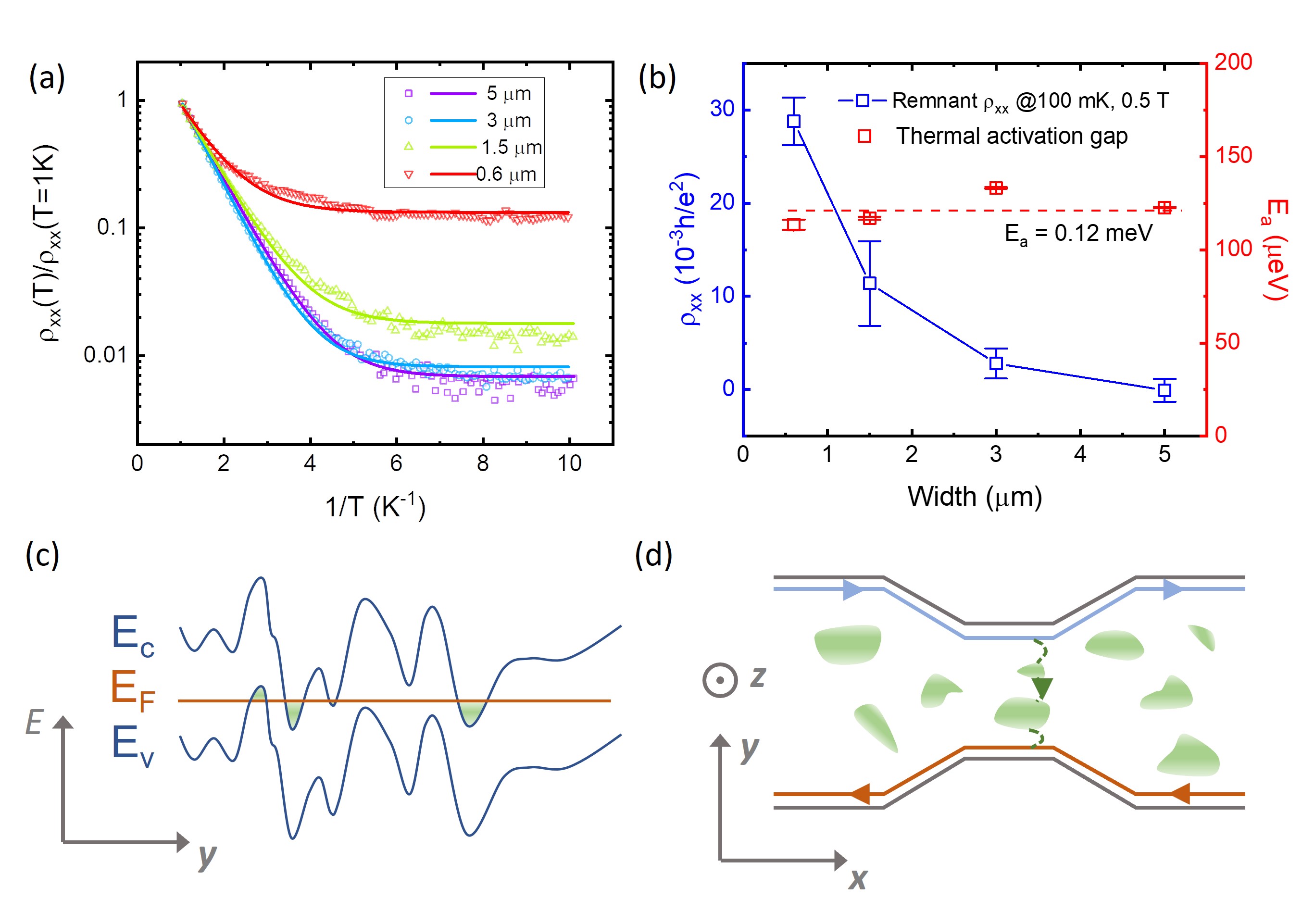}
\caption{\label{fig:epsart} Emerging back-scattering channels in narrow QAH devices. (a) Normalized temperature-dependent longitudinal resistance in an Arrhenius plot. 0.5 T external magnetic field was applied during cooling. The solid lines are fitted curves from Eq (1). (b) Remnant resistance (left axis) and thermal activation gap (right axis) in devices with different widths. Error bar in remnant resistance: s. d. from 20 data points. Error bar in activation gap: s. d. from curve fitting in (a). (c) Schematics showing energy fluctuations along the line cut across the channel due to inhomogeneity. (d) Back-scattering channels between opposite chiral edge modes in narrow QAH devices. Back-scattering processes are assisted by hopping through compressible 2D puddles. }
\end{figure}
The external magnetic field is biased at 400 mT which is above the coercive field at the base temperature. At high temperatures, the resistance drops exponentially with $1/T$ at the same decaying rate for all four sizes of devices, suggesting a bulk thermal activation gap independent of size dominates in this temperature regime. At low temperatures, the resistance saturates at a certain value, indicating a dissipative channel that originates from a temperature independent mechanism. The activation gap can be extracted by fitting the data with the following equation: \begin{equation}
\rho_{xx}=A \cdot exp(\frac{-E_a}{k_BT})+\rho_0
\end{equation}
 where $E_a$ is the activation energy, $k_B$ is the Boltzmann constant, $A$ is a device-dependent constant, and $\rho_0$ describes the remnant resistance contribution from the electron back scattering along percolating paths across two edges. The average activation gap $E_a$ is extracted to be 0.12 meV according to Eq. (1), and no obvious correlation with the geometric size can be drawn (see Fig. 3b). This implies the perturbation from quantum confinement effect in the \textit{y}-direction is negligible and the system can still be treated as a 2D QAH insulator. The extracted thermal activation gap is also close to the previously reported value \cite{Lei2021}. On the other hand, the remnant resistance $\rho_0$ increases monotonically with reduced size as plotted in Fig. 3b. This suggests a back-scattering mechanism is introduced despite topological protection when the width of the channel is reduced. Since the width of the narrowest device is still considerably larger than the Fermi wavelength, the direct overlapping of wavefunctions of chiral edge states on the opposite boundary is unlikely. However, the electrons can still be scattered to the opposite channel through percolative tunneling with the assistance of random compressible 2D puddles in the energy landscape (as shown in Fig. 3c and 3d). According to the Landauer formula, the conductance of a general quasi-1D system is related to the reflection coefficient R in the form of \cite{Buttiker1988}:
\begin{equation}
\sigma_{xx}=\frac{e^2}{h}\frac{1-R}{R}
\end{equation}
 From the remnant resistance, we can estimate a reflection coefficient $R$ of $0.21\%$ for the 0.6-$\mu$m-wide device and $0.01\%$ for the 5-$\mu$m-wide device due to cross channel back-scattering. A stronger size-dependent trend of remnant resistance was observed in another QAH sample with relatively low quality (see Supplementary Note 2). This can be explained by the finite-size percolation theory, where the self-similar renormalization transformation no longer holds when the size of the system is smaller than the coherence length. In this case, when approaching the quantum critical point from the subcritical phase, the probability of forming a percolating path will increase with decreasing sizes. This is reflected in the observation that the back-scattering is enhanced in narrower samples. We shall also mention that since the effective channel length also is reduced, back-scattering may also arise from intra-channel spin-flip scattering process in non-magnetic islands due to inhomogeneous magnetic doping, which can be treated as a 1D percolation problem. Both inter-channel back-scattering and intra-channel spin-scattering are phenomenologically described by the reflection coefficient $R$.

The QAH critical current was also measured in these devices with a differential resistance technique. A large DC current bias is sent through the device along with a small AC excitation down to 0.2 nA. Differential resistance was measured by a synchronized lock-in amplifier as a function of DC current bias in the configuration illustrated in the inset of Fig. 4b. All four devices show similar quantitative behavior (see Fig. 4a). 
\begin{figure}
\includegraphics[width=\linewidth]{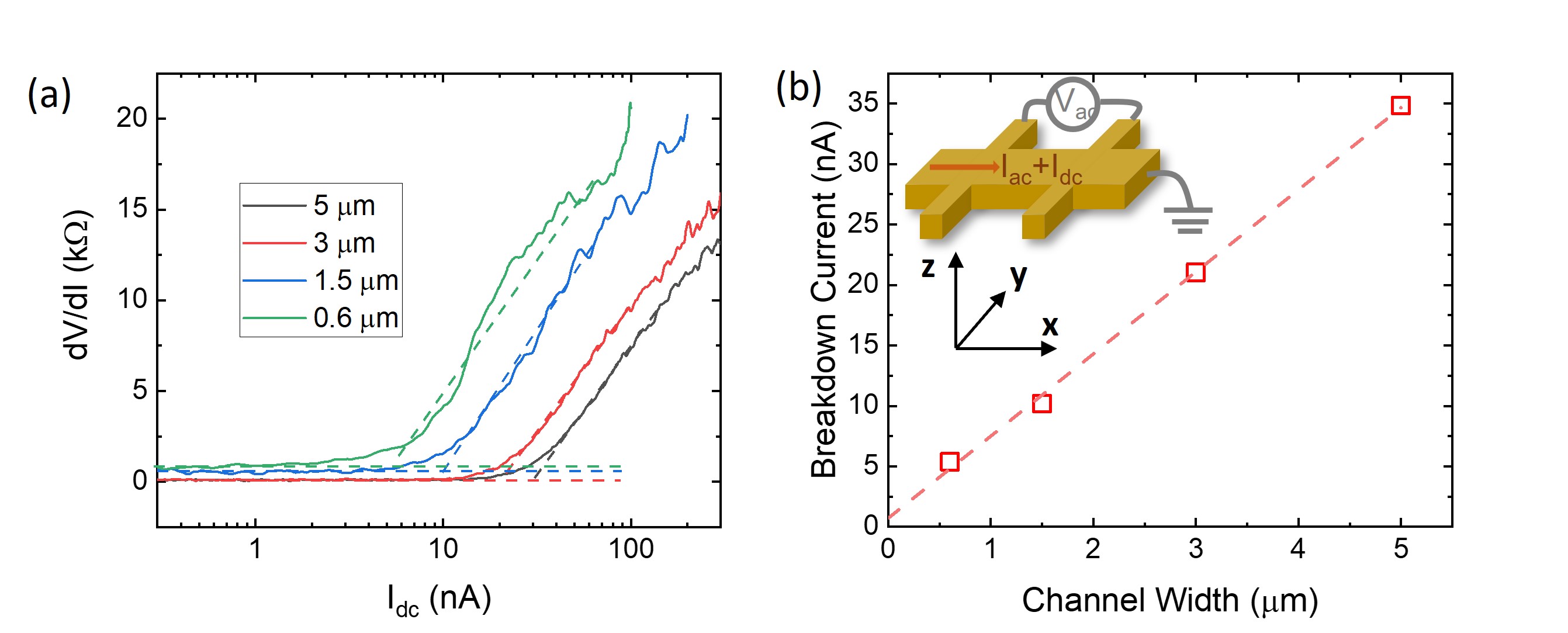}
\caption{\label{fig:epsart} Breakdown current in narrow QAH devices. (a) Differential resistance as a function of DC current bias. A low AC excitation of 0.2 nA is applied. Horizontal dashed lines are the average value of resistance in the linear region, and tilted dashed lines are plotted from linear fitting over a window of over 1 decade of $I_{dc}$. (b) Breakdown current versus channel width. The dashed line is the linear fitting of data. Inset, schematics of measurement setup for differential resistance.   }
\end{figure}
At low DC bias, the devices operate in a linear region where the remnant resistance is irrelevant to $I_{dc}$ but only determined by the previously mentioned back-scattering mechanism. Noted that all the other data throughout the paper were acquired under either 0.5 nA or 1 nA AC excitation, which falls into this linear region. Above breakdown current $I_{bd}$, the resistance begins to shoot up, as large current drives the system out of QAH states, which is attributed to the electric-field-driven percolation in the \textit{y}-direction \cite{Lippertz2021}. The breakdown current for all devices is extracted in Fig 4b, and it scales linearly with the channel width. This supports the argument of electric-field driven breakdown of QAH\cite{Lippertz2021}, according to the equation:
\begin{equation}
I_{x,bd} \frac{h}{e^2}=E_{y,bd}(W-2\lambda_F)
\end{equation}
 Here, $I_{x,bd}$ is the breakdown current in longitudinal direction, $E_{y,bd}$ is the QAH breakdown field in \textit{y}-direction, $W$ is the width of the channel and $\lambda_F$ is the Fermi wavelength of the chiral edge modes. Here we use reduced channel width $W-2\lambda_F$ instead of $W$, since the breakdown current should be proportional to the incompressible bulk region. It has been demonstrated in quantum Hall states, the Hall voltage profile in the \textit{y}-direction only drops in the incompressible bulk insulators, whereas the metallic edge channel has little voltage drop \cite{Weitz2000,WEIS2007}. The fact that the linear relationship between the breakdown current and the channel width extrapolates to near 0 on the \textit{x}-axis implies that the incompressible bulk states almost extend to the entire channel and the width of chiral edge states is negligible compared to the channel width. This confirms our previous argument that the chiral edge states in QAH only reside in close vicinity of the physical edge spanning the width of Fermi wavelength. 

Another prominent feature in narrow devices is the resistance fluctuations near the coercive field, which is associated with multi-domain behavior during topological phase transition. To quantitatively investigate the resistance fluctuations in the transition regimes, a smooth background was generated from raw data with a Savitzky–Golay filter over a window of 50 mT as shown in Fig. 5a. The resistance fluctuation $\Delta\rho_{xx}$ is then extracted by subtracting the raw data from the smooth background. One immediately notices the resistance fluctuations during magnetic switching are more pronounced in narrower devices (see Fig. 5b). Similar resistance fluctuations were also observed in narrow-channel QH devices during the plateau-to-plateau transition  \cite{Timp1987,Jain1988,Simmons1989,Main1994,Chang1988}.The fluctuations are suppressed at elevated temperatures as shown in Fig. 5c, confirming this is an mesoscopic transport behavior at low temperatures instead of measurement non-idealities. 
\begin{figure*}
\includegraphics[width=\linewidth]{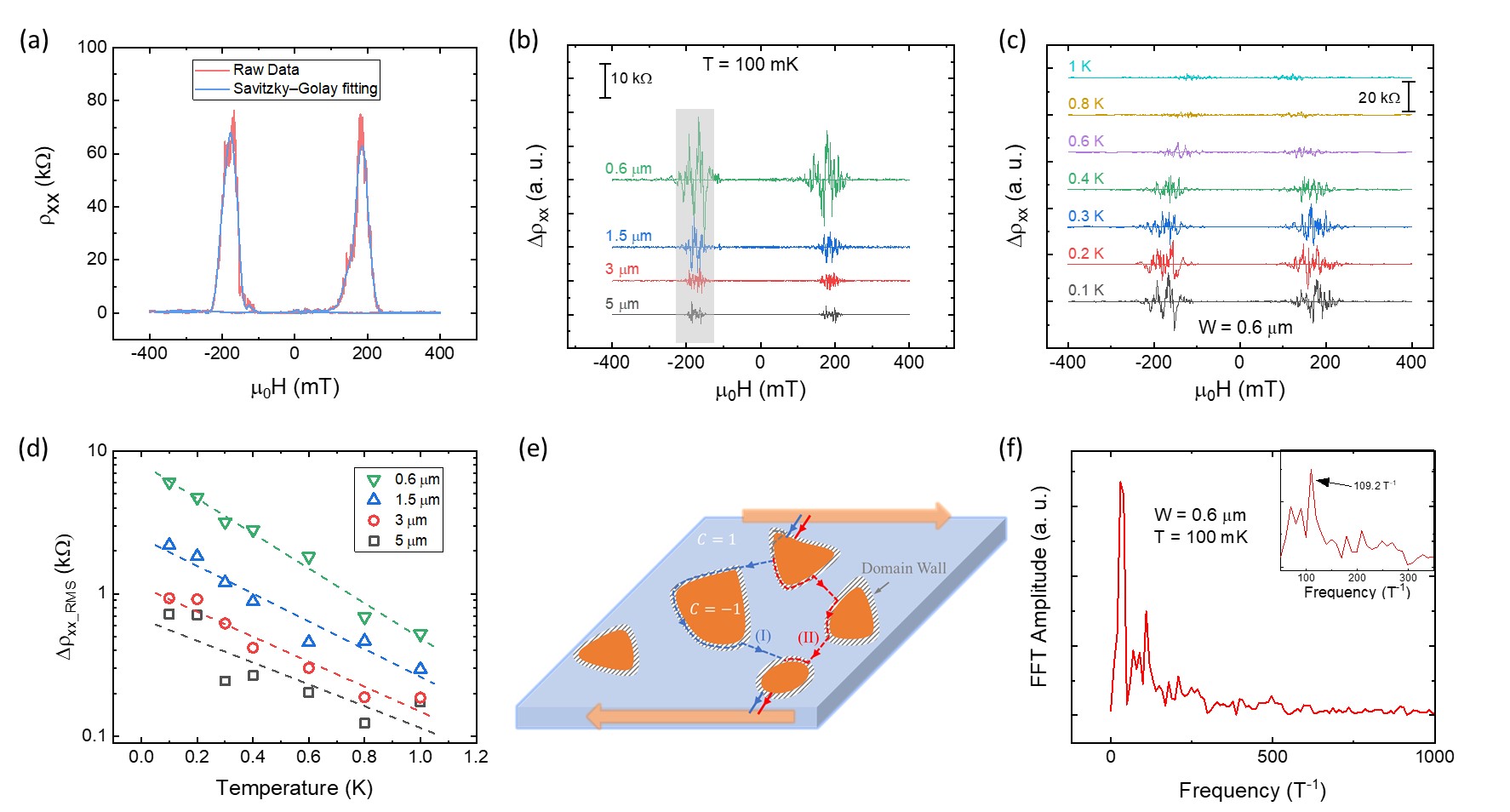}
\caption{\label{fig:epsart} Resistance fluctuations near the coercive field. (a) Extraction of resistance fluctuations by subtracting a smooth background with Savitzky–Golay filtering from raw data. (b) Resistance fluctuation $\Delta\rho_{xx}$ during field sweeping for devices with different widths measured at a base temperature of 100 mK.(c) Temperature dependence of resistance fluctuations measured in the 0.6-$\mu$m-wide device. The data in (b) and (c) is offset for clarity. (d) Fluctuation amplitudes decay with temperature. Resistance fluctuation amplitudes are quantitatively defined by the root-mean-square (RMS) value of the resistance fluctuation within a 100 mT window near the coercive field, as indicated by the grey region in Fig. 5a. Dashed lines are exponential decay fitting to the data. (e) Schematics of interfering paths (I) and (II) due to resonant tunneling between domains. (f) Fast Fourier transform of quasi-period resistance fluctuation. Inset: zoomed-in area near the dominant frequency at 109.2 $T^{-1}$.}
\end{figure*}

The fluctuations are quantitatively represented by the root-mean-square (RMS) value of $\Delta\rho_{xx}$ within a 100 mT window around the coercive field (as highlighted in the grey area in Fig. 5b) and plotted for all sizes at different temperatures in Fig 5d. $\Delta\rho_{xx\_RMS}$ decays exponentially with temperature in all devices, pointing to a behavior similar to the universal conductance fluctuation phenomena that is dictated by the phase coherence length which also follows the same temperature dependence. Here we semiclassically attribute the resistance fluctuations to the electron interference effect between intersecting tunneling paths in the percolation picture during topological phase transition, as depicted in Fig. 5e. Starting from the positive magnetic field, the MTI film is initially in $M_z=+1$ magnetization which corresponds to Chern number $C=1$. When the magnetic field is swept reversely approaching the coercive field, magnetic domains starts to flip into $M_z=-1$ ($C=-1$) domains. The domain walls between $C=1$ and $C=-1$ give rise to trivial compressible states. Electrons on the edge states can tunnel resonantly through domain walls in multiple paths. When the size of the sample is comparable or smaller than the electron phase coherence length $L_\phi$, the wavefunction of electrons in intersecting paths will interfere with each other in a similar way as in the Aharanov-Bohm interference effect. When the sample dimension is larger than the coherence length, electrons loses its phase information during inelastic scattering processes, and such fluctuations are no longer observable. Likewise, since the phase coherent length decreases with temperature, the fluctuations drop exponentially at higher temperatures. The similar mechanism also successfully explained resistance fluctuations in narrow QH devices during plateau-to-plateau transitions \cite{Jain1988}, with a minor difference being that in QH electrons resonantly tunnel through extended states rather than domain walls as in the QAH cases (see Fig. 5e). Since the extended states in QH films primarily arise due to lattice detects and substrate disorders, the fluctuation patterns are strongly reproducible \cite{Simmons1989}; whereas the fluctuation in narrow QAH films are less reproducible because the formation of domain walls is more subject to domain dynamics (see Supplementary Note 4). Finally, the quasi-period resistance fluctuation gives a good measure of the average domain size, as whether the interference is destructive or constructive depends on the magnetic flux quanta enclosed within the interference paths around domains. The Fourier frequency spectrum of resistance fluctuation near the coercive field in the W=0.6 $\mu$m device is shown in Fig. 5f. Here the first large low frequency peak should not be interpreted as a fluctuation frequency, as its period ($\sim$33.6 mT) is comparable to the observation window (100 mT). This peak is due to the boundary condition imposed when subtracting a smooth background. A dominant frequency associated with quasi-period fluctuations is located at 109.2 $T^{-1}$, which corresponds to a quasi-period of $\sim$9.2 mT. This primary period can be translated into an average radius of domain size of 280 nm following the equation:
\begin{equation}
\Delta B\cdot\pi R^2=\Phi_0
\end{equation}

 This estimated characteristic size of superparamagnetic domain size is comparable to the value (average diameter of 175 nm at 300 mK \cite{Lachman2017}) obtained from direct imaging using scanning nano-SQUID imaging techniques  \cite{Lachman2017,Wang2018,Lachman2021}.
 
\section{\label{sec:level1}Conclusion}
In this work, we presented a detailed study on the transport behavior of QAH insulators in the sub-micron size regime. QAH signatures, including vanished longitudinal resistance and quantized Hall resistance, are preserved in narrow-channel devices down to 600 nm. We argue that QAH systems are more suitable than QH systems for implementing nano-scale quantum devices since their chiral edge modes are tightly bounded to the physical boundary within the order of Fermi wavelength. We confirmed the compressible chiral edge states have negligible width by measuring width-dependent breakdown current. An emerging cross-edge back-scattering channel is observed in sub-micron QAH devices, which is qualitatively explained by finite-size percolation theory. Resistance fluctuations near the coercive field was observed in devices narrower than the phase coherence length $L_\phi$, which is attributed to the coherent interference between intersecting paths around emerging domain walls. The quasi-period of the oscillations gives an estimated average domain radius of 280 nm, which matches well with the value obtained from direct imaging techniques. Our work provided a comprehensive understanding and insights of the mesoscopic transport of the QAH effect both in quantized states and during topological phase transitions.

\begin{acknowledgments}
This work was supported by the NSF under Grants No. 1936383 and No. 2040737, the U.S. Army Research Office MURI program under Grants No. W911NF-20-2-0166 and No. W911NF-16-1-0472. C.E. is an employee of  Fibertek, Inc. and performs in support of Contract No.W15P7T19D0038, Delivery Order W911-QX-20-F-0023 . The views expressed are those of the authors and do not reflect the official policy or position of the Department of Defense or the US government. The identification of any commercial product or tradename does not imply endorsement or recommendation by Fibertek Inc. G. Q. would like to thank Dr. Z Wan and Dr. Q. Qian for valuable discussions on mesoscopic transport in QH systems.
\end{acknowledgments}

\bibliography{apssamp}

\end{document}